\documentclass[fleqn,10pt]{wlscirep}
\usepackage{subfigure}
\usepackage{algorithm}
\usepackage{algpseudocode}
\title{Capacity limit for faster-than-Nyquist non-orthogonal frequency-division multiplexing signaling}

\author[1, 2]{Ji Zhou}
\author[1,*]{Yaojun Qiao}
\author[3]{Zhanyu Yang}
\author[2]{Qixiang Cheng}
\author[4]{Qi Wang}
\author[1]{Mengqi Guo}
\author[1]{Xizi Tang}
\affil[1]{State Key Laboratory of Information Photonics and Optical Communications, School of Information and Communication Engineering, Beijing University of Posts and Telecommunications (BUPT), Beijing 100876, China}
\affil[2]{Department of Electrical Engineering, Columbia University, New York 10026, USA}
\affil[3]{Department of Electrical and Computer Engineering, University of Virginia, Virginia 22904, USA}
\affil[4]{School of Electronics and Computer Science, University of Southampton, Southampton SO17 1BJ, U.K.}

\affil[*]{qiao@bupt.edu.cn}

\begin{abstract}
Faster-than-Nyquist (FTN) signal achieves higher spectral efficiency and capacity compared to Nyquist signal due to its smaller pulse interval or narrower subcarrier spacing. Shannon limit typically defines the upper-limit capacity of Nyquist signal. To the best of our knowledge, the mathematical expression for the capacity limit of FTN non-orthogonal frequency-division multiplexing (NOFDM) signal is first demonstrated in this paper. The mathematical expression shows that FTN NOFDM signal has the potential to achieve a higher capacity limit compared to Nyquist signal. In this paper, we demonstrate the principle of FTN NOFDM by taking fractional cosine transform-based NOFDM (FrCT-NOFDM) for instance. FrCT-NOFDM is first proposed and implemented by both simulation and experiment. When the bandwidth compression factor $\alpha$ is set to $0.8$ in FrCT-NOFDM, the subcarrier spacing is equal to $40\%$ of the symbol rate per subcarrier, thus the transmission rate is about $25\%$ faster than Nyquist rate. FTN NOFDM with higher capacity would be promising in the future communication systems, especially in the bandwidth-limited applications.
\end{abstract}

\begin{document}

%\begin{linenumbers}

\flushbottom

\maketitle
\thispagestyle{empty}

%\noindent Please note: Abbreviations should be introduced at the first mention in the main text – no abbreviations lists. Suggested structure of main text (not enforced) is provided below.

\section*{Introduction}
In the 1940s, an upper limit of communication capacity was proposed for Nyquist signal in an additive white Gaussian noise (AWGN) channel, now known as the Shannon limit \cite{Shannon:1948TBSTJ, Shannon:1949PIRE}. Initially, the communication capacity is enough for the traditional data and voice services. However, with the exponential growth of data traffic due to bandwidth-intensive applications such as high definition TV and mobile video, the communication capacity gradually approaches Shannon limit nowadays \cite{Essiambre:2010JLT, Essiambre:2012IEEE, Liu:2013NP}. Increasing the spectral efficiency is a key challenge to meet the increasing demand for higher capacity over communication channels. Faster-than-Nyquist (FTN) signal was first proposed by Mazo in 1970s to improve the spectral efficiency \cite{Mazo:1975TBSTJ}. As the name implies, FTN signal can achieve a symbol rate faster than Nyquist rate. Therefore, it has been widely investigated in high-capacity wireless and optical communications \cite{ Anderson:2013IEEE, Andrews:2014JSAC, Zhang:2015SR, Kaneda:2014OECC, Sato:2014ECOC, Zhang:2015OE, Igarashi:2014OFC}.

The reported FTN scheme can be generally categorized into two types, one is compressing the duration between the adjacent pulses in time domain\cite{Mazo:1975TBSTJ}, and the other one is compressing the baseband bandwidth in frequency domain\cite{Zhang:2015OE, Igarashi:2014OFC, Zhou:2016OL}. In time domain, FTN binary sinc-pulse signal was proposed in 1975 by accelerating the pulse with time acceleration factor $\tau$. The accelerated pulses are no longer orthogonal, thus FTN binary sinc-pulse signal is a kind of FTN non-orthogonal time-division multiplexing (NOTDM) signal. When $\tau$ is set to $0.8$, known as the Mazo limit, $25\%$ more bits can be carried in the same bandwidth and there is not obvious deterioration in bit error rate (BER) performance by using the trellis decoding to effectively compensate the inter-symbol interference (ISI) \cite{Anderson:2013IEEE}. 
In frequency domain, the duobinary-pulse shaping filter is employed to compress the baseband bandwidth of single-carrier signal \cite{Zhang:2015OE, Igarashi:2014OFC} or the subcarrier spacing of multi-carrier signal is compressed to obtain narrower baseband bandwidth \cite{Zhou:2016OL}. In this paper, we focus on the multi-carrier FTN signal. Non-orthogonal frequency-division multiplexing (NOFDM) signal has been realized by further compressing the subcarrier spacing compared to OFDM signal \cite{Nopchinda:2016PTL, Darwazeh:2014PTL, Darwazeh:2013CL}. We found that only when the subcarrier spacing is less than half of the symbol rate per subcarrier, NOFDM signal can achieve a symbol rate faster than Nyquist rate as we demonstrated in our recent work \cite{Zhou:2016OL}.

FTN signal achieves higher spectral efficiency and capacity compared to Nyquist signal due to its smaller pulse interval or narrower subcarrier spacing. Shannon limit typically defines the upper-limit capacity of Nyquist signal. In this paper, we first give the mathematical expression for the capacity limit of FTN NOFDM signal. The mathematical expression shows that FTN signal has the potential to  achieve a higher capacity limit compared to Nyquist signal. In this paper, we demonstrate the principle of FTN NOFDM by taking fractional cosine transform-based NOFDM (FrCT-NOFDM) for instance. FrCT-NOFDM is first proposed and implemented by both simulation and experiment. When the bandwidth compression factor $\alpha$ is set to $0.8$, the subcarrier spacing is equal to $40\%$ of the symbol rate per subcarrier, thus transmission rate is about $25\%$ faster than Nyquist rate.

\section*{Results}

\begin{figure}[!tb]
\centering
\includegraphics[width=6in]{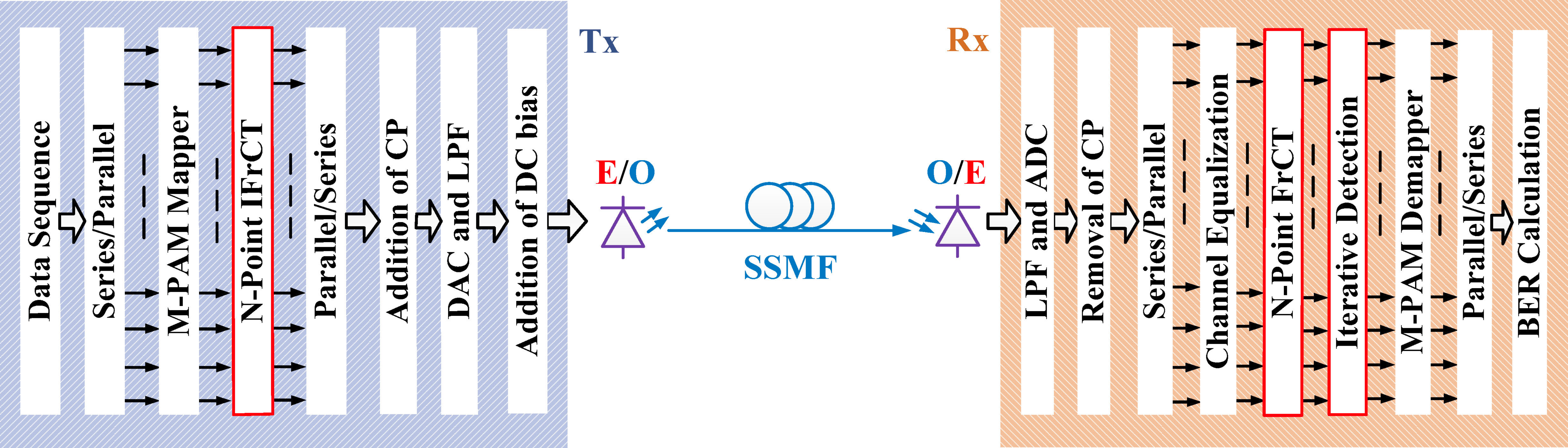}
\caption{Block diagram of FrCT-NOFDM for optical transmission system.}
\label{fig:1}
\end{figure}

In this paper, we demonstrate the principle of FTN NOFDM by taking FrCT-NOFDM for instance. A block diagram of the FrCT-NOFDM system is depicted in Figure \ref{fig:1}. Different from traditional OFDM system, the inverse FrCT (IFrCT)/FrCT algorithm is employed to implement the multiplexing/demultiplexing processing. The $N$-order IFrCT and FrCT are defined as
\begin{equation}\label{eq:eq1}
x_{n}=\sqrt{\frac{2}{N}}\sum_{k=0}^{N-1}W_{k}X_{k}\text{cos}\left(\frac{\pi \alpha (2n+1)k}{2N}\right),
\end{equation}
\begin{equation}\label{eq:eq2}
X_{k}=\sqrt{\frac{2}{N}}W_{k}\sum_{n=0}^{N-1}x_{n}\text{cos}\left(\frac{\pi \alpha (2n+1)k}{2N}\right)
\end{equation}
where $0\leq n\leq N-1$, $0\leq k\leq N-1$,
\begin{equation}\label{eq:eq3}
W_{k} =
\begin{cases}
\frac{1}{\sqrt{2}}, &\text{$k=0$}\\
1, &\text{$k=1,~2,~\cdots,~N-1$}
\end{cases}
\end{equation}
and $\alpha$ is the bandwidth compression factor. The $\alpha$ less than $1$ determines the level of the bandwidth compression. When $\alpha$ is equal to $1$, Equation \ref{eq:eq2} is the Type-II discrete cosine transform (DCT) in which the matrix is orthogonal. The Type-II DCT is probably the most commonly used form, and is often simply referred to as ``the DCT'' \cite{Ahmed:1974TC}. DCT-based OFDM (DCT-OFDM) signal can be generated by Type-II DCT \cite{Zhou:2015JLT,Zhou:2016JLT}. In DCT-OFDM, the subcarrier spacing is equal to half of the symbol rate per subcarrier, which is half of the subcarrier spacing in discrete Fourier transform-based OFDM (DFT-OFDM). Therefore, the subcarrier spacing of DCT-OFDM is equal to $1/2T$ where $T$ denotes the time duration of one DCT-OFDM symbol \cite{Peng:2006TC,Zhao:2010OFC}. Due to the compression of subcarrier spacing, the subcarrier spacing of FrCT-NOFDM should be smaller than $1/2T$.

\subsection*{Spectral efficiency and capacity limit}
\begin{figure}[!t]
\centering
\includegraphics[width=4in]{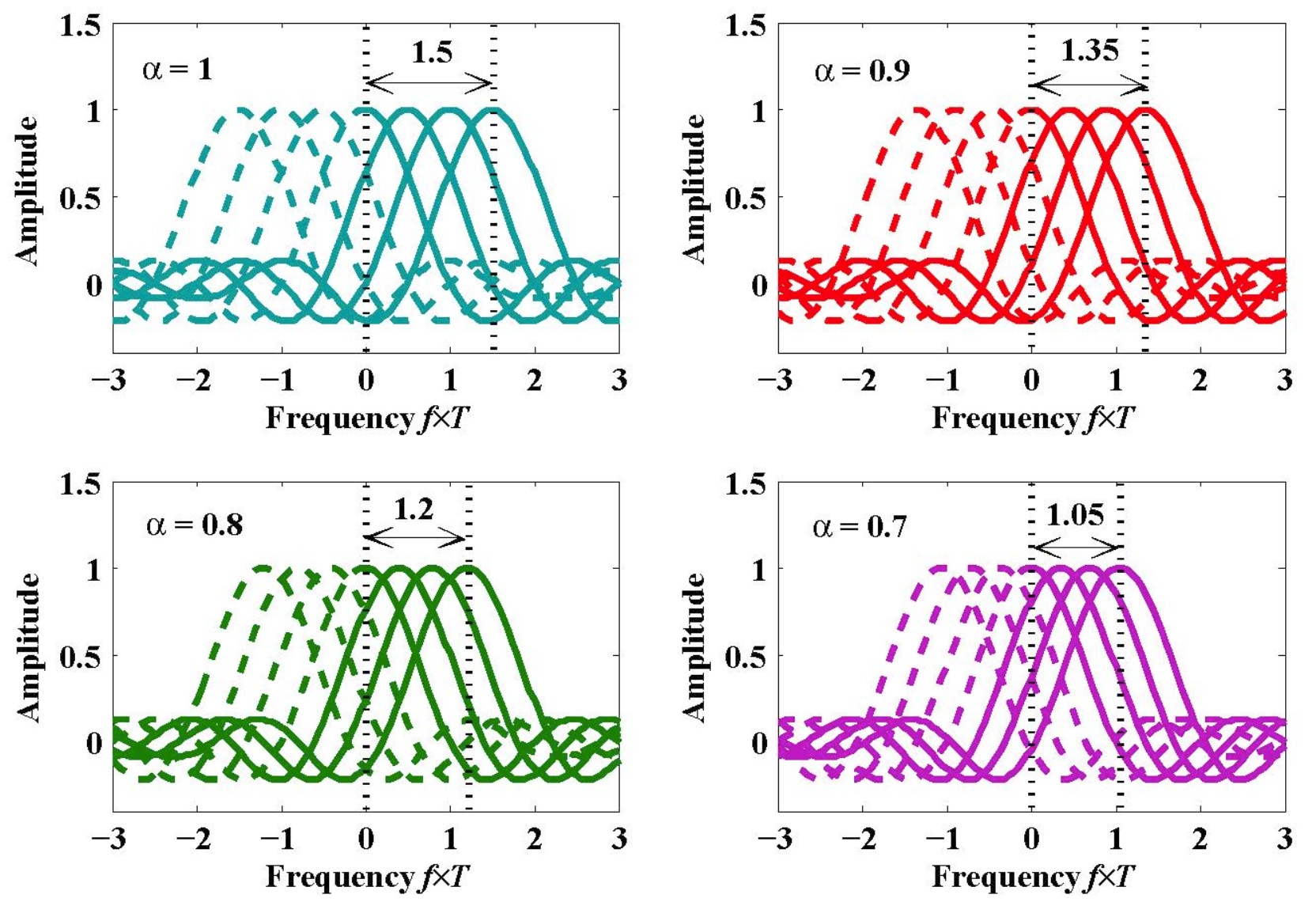}
\caption{Sketched spectra of DCT-OFDM (i.e., $\alpha = 1$) and FrCT-NOFDM.}
\label{fig:2}
\end{figure}

Figure \ref{fig:2} shows the sketched spectra of DCT-OFDM (i.e., $\alpha = 1$) and FrCT-NOFDM when the subcarrier number is set to $4$. The subcarrier spacing of FrCT-NOFDM is equal to $\alpha/2T$ where $T$ denotes the time duration of one FrCT-NOFDM symbol. All the subcarriers locate in the positive frequency domain. Meanwhile, their images fall into the negative frequency domain. When the subcarrier number is large enough, the baseband bandwidth of FrCT-NOFDM can be calculated by
\begin{equation}\label{eq:eq4}
\text{Baseband bandwidth} = \frac{(N-1)\times \alpha}{2\times T}+\frac{1}{T} \approx \frac{N\times \alpha}{2\times T}
\end{equation}
where $N$ denotes the subcarrier number.

In general, Nyquist frequency is equal to half of the sample rate, which is equal to $N/2T$. When $N$ is large enough, the baseband bandwidth of DCT-OFDM (i.e., $\alpha = 1$) is almost the same with Nyquist frequency and the baseband bandwidth of FrCT-NOFDM (i.e., $\alpha < 1$) is smaller than Nyquist frequency. It is worth noting that Nyquist frequency should not be confused with Nyquist rate. Nyquist rate is defined as the twice of the baseband bandwidth,
\begin{equation} \label{eq:eq5}
R_{N} = \frac{N\times \alpha}{T}.
\end{equation}
As we know, the symbol rate of FrCT-NOFDM is equal to the sample rate when all the subcarriers are valid,
\begin{equation} \label{eq:eq6}
R_{S} = \frac{N}{T},
\end{equation}
thus the symbol rate of FrCT-NOFDM is faster than Nyquist rate. FrCT-NOFDM can be considered as a kind of FTN signal. When $\alpha$ is set to $0.8$, the subcarrier spacing is equal to $40\%$ of symbol rate per subcarrier. Twenty percent baseband bandwidth saving can be obtained and the transmission rate is about $25\%$ faster than the Nyquist rate.

Afterwards, we discuss the capacity limit for FrCT-NOFDM signal. The channel is in the presence of AWGN. The well-known Shannon limit for Nyquist signal can be calculated by
\begin{equation}\label{eq:eq7}
C \leq  Wlog_{2}\left(1+\frac{P_S}{P_N}\right)
\end{equation}
where $W$ is the signal bandwidth, $P_S$ is the signal power and $P_N$ is the noise power. This equation depicts the upper-limit capacity of Nyquist signal\cite{Shannon:1949PIRE}.

To obtain the capacity limit of FrCT-NOFDM signal, we will apply the analysis method for single-carrier signal in Shannon's papers \cite{Shannon:1948TBSTJ, Shannon:1949PIRE} to the multi-carrier signal. However, there is great difference between single-carrier and multi-carrier signal. The capacity limit of the single-carrier signal can be carried out in the time domain, but the capacity limit of multi-carrier signal should be derived out in the frequency domain. The derivation of capacity limit needs to employ the geometric methods \cite{Sommerville:1929}. In the following, we will briefly give the derivation of the capacity limit for FrCT-NOFDM. This derivation is similar to that in Shannon's papers.

In the frequency domain, there are $W/(subcarrier~spacing)$ independent amplitudes where $W$ is the baseband bandwidth and the subcarrier~spacing is equal to $\alpha/2T$. Therefore, the number of independent amplitudes is equal to $2TW/\alpha$, which determines the dimensions of signals. For large $W$, the perturbation caused by noise can be considered as some points near the surface of a sphere with radius $\sqrt{2TWP_N/\alpha}$ centered at the signal point. The power of received signal is $P_S+P_N$. Similar to the perturbation, the received signal can be considered as some points whose positions are on the surface of a sphere with radius $\sqrt{2TW(P_S+P_N)/\alpha}$. The number of the distinguishable signals is no more than the volume of the sphere with radius $\sqrt{2TW(P_S+P_N)/\alpha}$ divided by the volume of the sphere with radius $\sqrt{2TWP_N/\alpha}$. The volume of an $n$-dimensional sphere of radius $r$ can be calculated by \cite{Sommerville:1929}
\begin{equation}\label{eq:eq8}
V = \frac{\pi^{n/2}}{\Gamma\big(n/2+1\big)}r^n.
\end{equation}
where $\Gamma\big(n/2+1\big) = \int_{0}^{\infty} e^{-t}\times t^{n/2} dt$. Therefore, the upper limit for the number of distinguishable signals is given by
\begin{equation}\label{eq:eq9}
M \leq \left(\sqrt{\frac{P_S+P_N}{P_N}}\right)^{2WT/\alpha}.
\end{equation}
Consequently, the capacity limit of FrCT-NOFDM signal can be bounded by
\begin{equation}\label{eq:eq10}
C =\frac{log_{2}M}{T} \leq \frac{1}{\alpha}\times  Wlog_{2}\left(1+\frac{P_S}{P_N}\right)
\end{equation}

\begin{figure}[!t]
\centering
\includegraphics[width=3in]{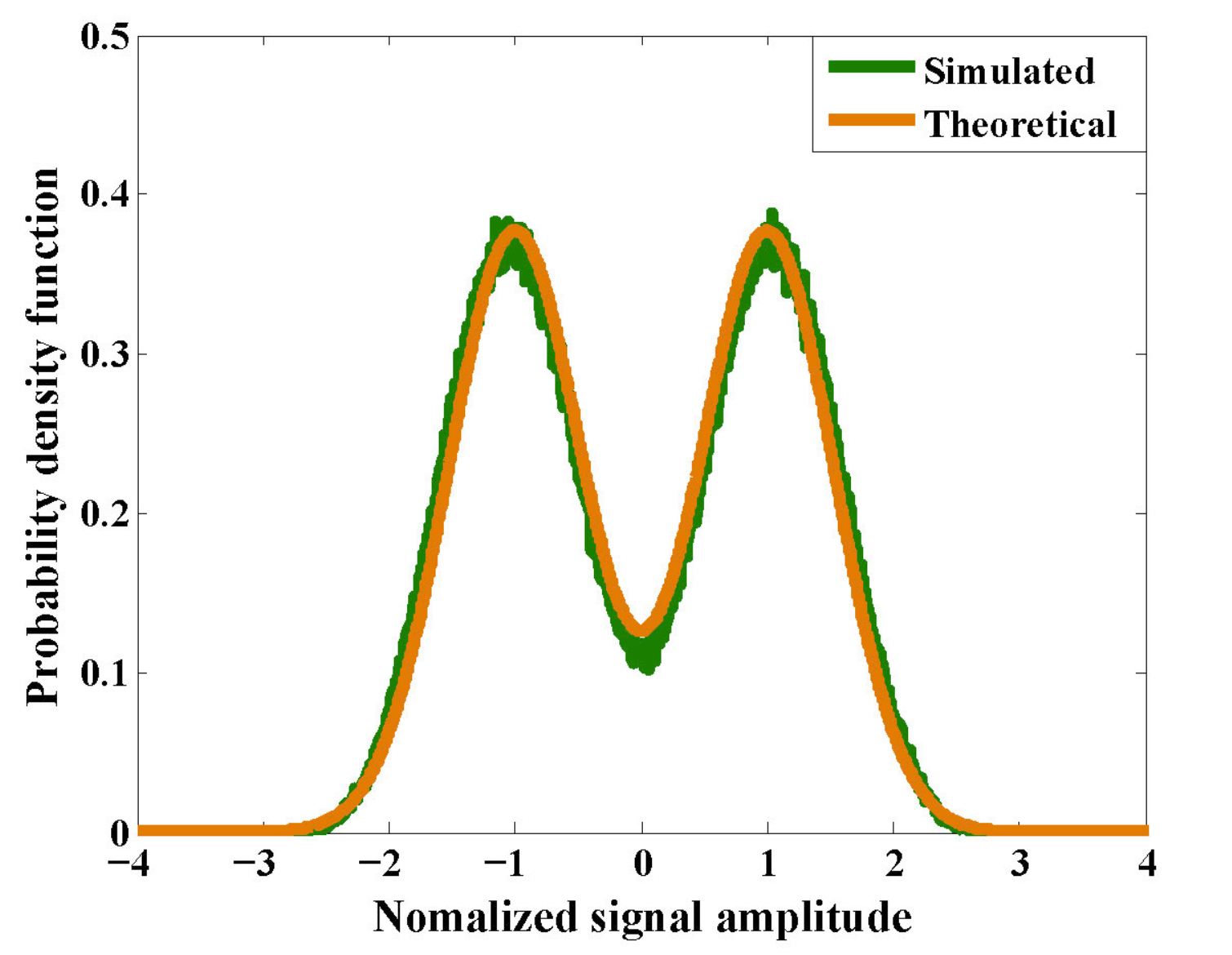}
\caption{The probability density function (PDF) of 2-PAM with inter-carrier interference (ICI) in FrCT-NOFDM with $\alpha$ of 0.8.}
\label{fig:3}
\end{figure}

It is worth noting that, in FrCT-NOFDM signal, inter-carrier interference (ICI) is a serious problem that should not be neglected because the subcarriers are no longer orthogonal. Figure \ref{fig:3} shows the probability density function (PDF) of ICI in FrCT-NOFDM with $\alpha$ of 0.8. The simulated curve denotes the PDF of the 2-PAM with ICI in FrCT-NOFDM with $\alpha$ of 0.8. The theoretical curve depicts the PDF of the 2-PAM with Gaussian noise, which can be defined as
\begin{equation}\label{eq:eq11}
PDF_{Theoretical}(x) = \frac{1}{2\sqrt{2\pi}\sigma}\left\{e^{-\frac{(x+1)^2}{2\sigma^2}}+e^{-\frac{(x-1)^2}{2\sigma^2}}\right\}
\end{equation}
where $\sigma^2$ is the variance of the Gaussian noise. The simulated curve agrees well with the theoretical curve. Therefore, the distribution of ICI is approximately Gaussian. Moreover, the ICI is independent of the Gaussian noise in AWGN channel. Considering both ICI and Gaussian noise in AWGN channel, the capacity limit of FrCT-NOFDM signal can be rewritten as
\begin{equation}\label{eq:eq12}
C \leq \frac{1}{\alpha}\times  Wlog_{2}\left(1+\frac{P_S}{P_N+P_{ICI}}\right)
\end{equation}
where $P_{ICI}$ is the power of ICI, which degrades the BER performance and thus decreases the capacity of FrCT-NOFDM signal. Therefore, the capacity of FrCT-NOFDM signal is likely to approach the capacity limit in Equation \ref{eq:eq10} only when ICI has been effectively eliminated. As verified in literatures \cite{Mazo:1975TBSTJ, Anderson:2013IEEE}, when the $\alpha$ is set to the value between $1$ and $0.8$, there is not obvious deterioration in BER performance by employing the optimal detection to eliminate the interference. The capacity limit of FrCT-NOFDM signal is potentially higher than that of the Nyquist signal when the $\alpha$ is set to the value between 1 and 0.8. How to eliminate the ICI is crucial for FrCT-NOFDM system. Afterwards, we will demonstrate the simulation and experiment to investigate the performance of FrCT-NOFDM system and verify the feasibility of the ICI cancellation algorithm. 

\begin{figure}[!t]
\centering
\includegraphics[width=4in]{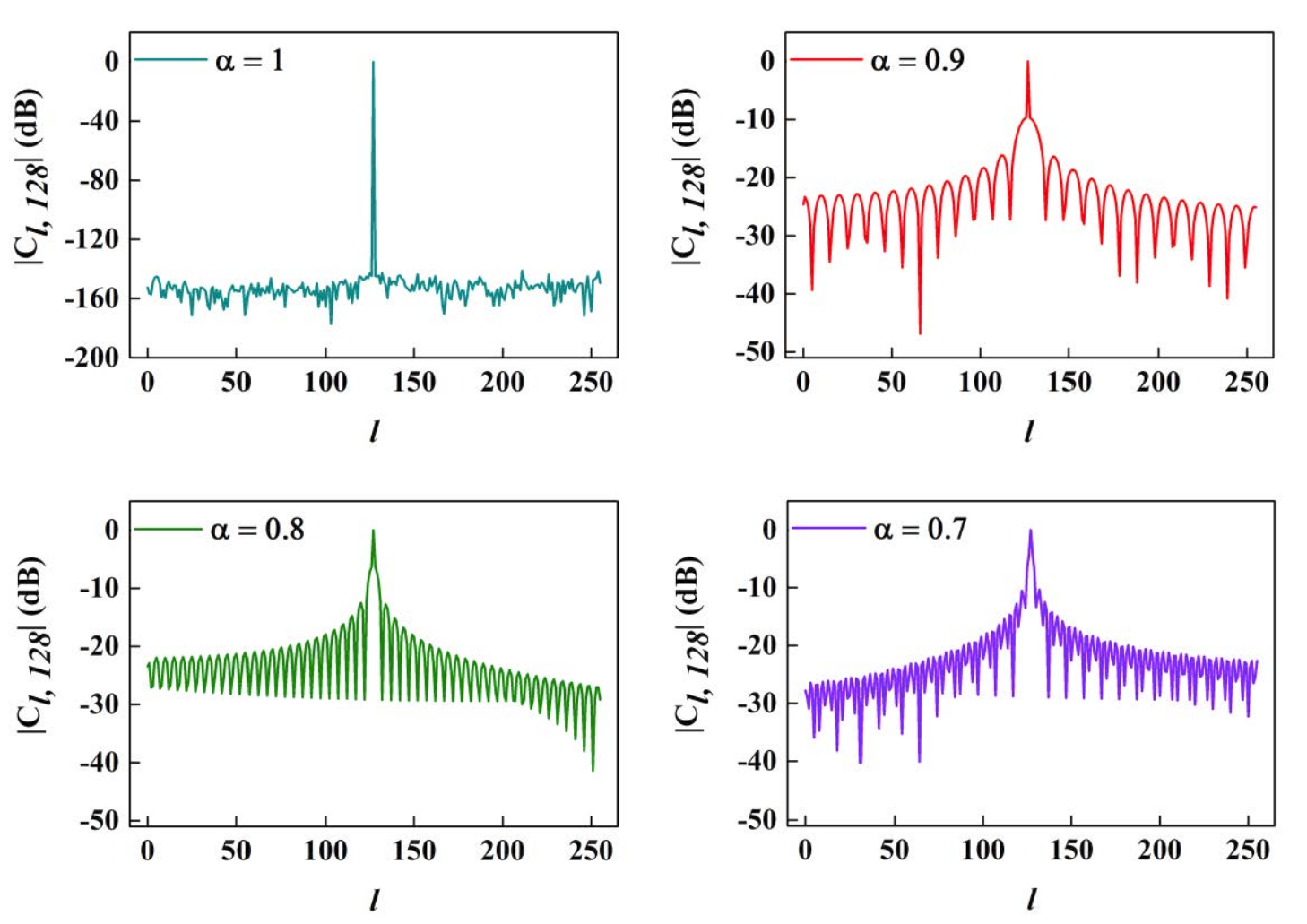}
\caption{$|C_{l, 128}|$ versus $l$ for DCT-OFDM (i.e., $\alpha = 1$) and FrCT-NOFDM.}
\label{fig:4}
\end{figure}

\subsection*{Simulation and theoretical analysis}
In this section, we will introduce an ICI cancellation algorithm based on iterative detection (ID) for FrCT-NOFDM. The simulations are demonstrated to verify the feasibility of FrCT-NOFDM. In the simulation system, the AWGN channel is employed. The number of subcarriers is set to $256$ and the number of FrCT-NOFDM symbols is set to $4096$. $E_{b}/N_{0}$ denotes the ratio between the energy per bit and the single-sided noise power spectral density.

An $N\times N$ correlation matrix $\textbf{C}$ is defined to study the ICI of FrCT-NOFDM, in which the interference between subcarriers $l$ and $m$ is represented by $C_{l,~m}$, the cross-correlation value. When $l$ equals to $m$, $C_{l,~m}$ represents the auto-correlation value for each of the subcarriers. $C_{l,~m}$ can be calculated by
\begin{equation}\label{eq:eq13}
C_{l,~m} = \frac{2}{N}\sum_{n=0}^{N-1}W_{l}\text{cos}\left(\frac{\alpha \pi l(2n+1)}{2N}\right)\cdot W_{m}\text{cos}\left(\frac{\alpha \pi (2n+1)m}{2N}\right).
\end{equation}

Figure \ref{fig:4} presents the $|C_{l,128}|$ as a function of $l$ for DCT-OFDM (i.e., $\alpha = 1$) and FrCT-NOFDM. When $\alpha$ is set to $1$, the auto-correlation value for the $128^{th}$ subcarrier is $1$. In the same time, the cross-correlation values are almost equal to $0$. Therefore, when $\alpha$ is set to $1$, $\textbf{C}$ is an identity matrix and the output of DCT is the OFDM signal. To achieve smaller subcarrier spacing, $\alpha$ can be set to be smaller than $1$. However, when $\alpha$ is less than $1$, the cross-correlation values are no longer equal to 0 and the orthogonality is destroyed. As shown in Figure \ref{fig:4}, the interference caused by the adjacent subcarriers is larger than that caused by farside subcarriers. The ICI increases with the decrease of $\alpha$. Thus, BER performance of FrCT-NOFDM degrades with the decrease of $\alpha$. When the subcarrier spacing is equal to the half of symbol rate per subcarrier, the subcarriers are orthogonal in DCT-OFDM, but the counterparts are no longer orthogonal in fractional Hartley transform NOFDM (FrHT-NOFDM) \cite{Zhou:2016OL} or fractional Fourier transform-based NOFDM (FrFT-NOFDM) \cite{Darwazeh:2013CL}. Therefore, under the same subcarrier spacing, the ICI in FrCT-NOFDM should be smaller than that in FrHT-NOFDM or FrFT-NOFDM. 

Recently, ID algorithm has been researched to eliminate the ICI for FrFT-NOFDM \cite{Darwazeh:2013CL, Darwazeh:2014PTL, Nopchinda:2016PTL}. FrFT-NOFDM is modulated using two-dimensional constellation (e.g., $M$-QAM), while, as Figure \ref{fig:1} depicts, FrCT-NOFDM is modulated using one-dimensional constellation (e.g., $M$-PAM). The ID algorithm for FrCT-NOFDM with one-dimensional constellation can been given by
\begin{equation}\label{eq:eq14}
\textbf{S}_{i} =\textbf{R} - (\textbf{C}-\textbf{e})\textbf{S}_{i-1}
\end{equation}
where $\textbf{R}$ is an $N$-dimensional vector of the received symbol demodulated by FrCT, $\textbf{S}_i$ is an $N$-dimensional vector of the recovered symbol after $i^{th}$ iteration, and $\textbf{e}$ is an $N\times N$ identity matrix.

\begin{figure}[!t]
\centering
\includegraphics[width=3in]{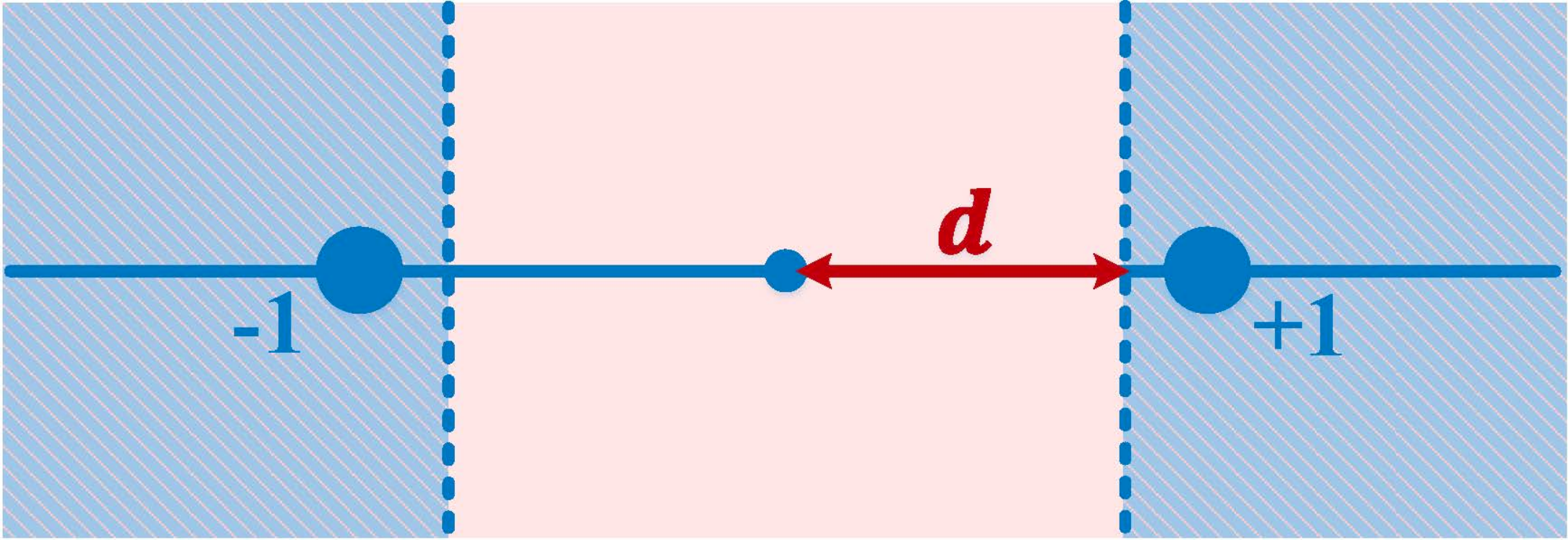}
\caption{Mapping strategy for $2$-PAM constellation.}
\label{fig:5}
\end{figure}

\begin{figure}[!t]
\centering
\includegraphics[width=3in]{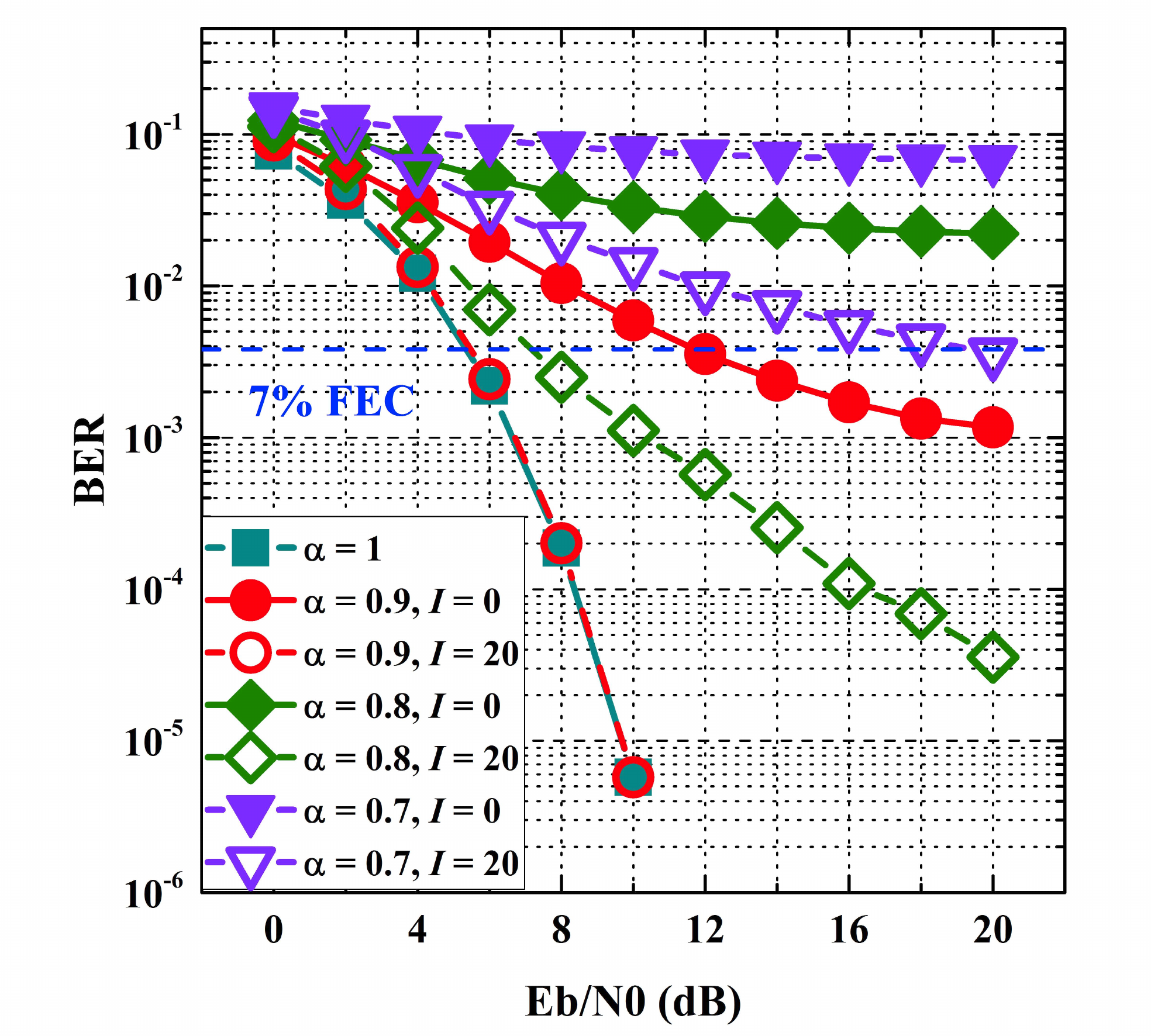}
\caption{BER against $E_{b}/N_{0}$ for DCT-OFDM (i.e., $\alpha = 1$) and FrCT-NOFDM.}\label{fig:6}
\end{figure}

The mapping strategy of 2-PAM constellation is outlined by Fig. \ref{fig:5}. Only points that fall in the blue area can be mapped to the corresponding constellation points. The points that fall in the uncertainty interval, which is from $-d$ to $d$ (i.e., the pink area), are unchanged. The $d$ is equal to $1-i/I$ where $i$ denotes the $i^{th}$ iteration and $I$ is the total iterative number. In each iteration, the operation of Equation (\ref{eq:eq13}) is implemented to eliminate the ICI. As a result, ICI is gradually reduced after each iteration and thus the pink area can be decreased. The algorithm terminates when $d$ is equal to zero. Since the decision for one-dimensional constellation is more simple than that for two-dimensional constellation, ID algorithm of FrCT-NOFDM has lower computational complexity than that of FrFT-NOFDM.

Figure \ref{fig:6} depicts the BER against $E_{b}/N_{0}$ for DCT-OFDM (i.e., $\alpha = 1$) and FrCT-NOFDM. The modulated constellation employs $2$-PAM. $I$ denotes the iterative number of ID algorithm. When $\alpha$ is less than $1$, the BER performance of FrCT-NOFDM without ID algorithm (i.e., $I$ is set to $0$) is seriously influenced by ICI. When $I$ is set to $20$, FrCT-NOFDM with $\alpha$ of $0.9$ has the same BER performance compared to the DCT-OFDM. This is because that ID algorithm can effectively eliminate the ICI in FrCT-NOFDM with $\alpha$ of $0.9$. When $\alpha$ is set to $0.8$ or $0.7$, the BER performance is still influenced by the residual ICI although the ID algorithm is employed. At the $7\%$ forward error correction (FEC) limit, the required $E_{b}/N_{0}$ for FrCT-NOFDM with $\alpha$ of $0.8$ is about $2$ dB higher than that for DCT-OFDM. If the FEC coding technique is employed, FrCT-NOFDM with $\alpha$ of $0.8$ would have almost the same performance with DCT-OFDM. When $\alpha$ is set to $0.7$, the BER of FrCT-NOFDM only achieves the $7\%$ FEC limit. The BER performance degrades with the decrease of $\alpha$ due to the increase of residual ICI.

\begin{figure}[!t]
\centering
\includegraphics[width=3in]{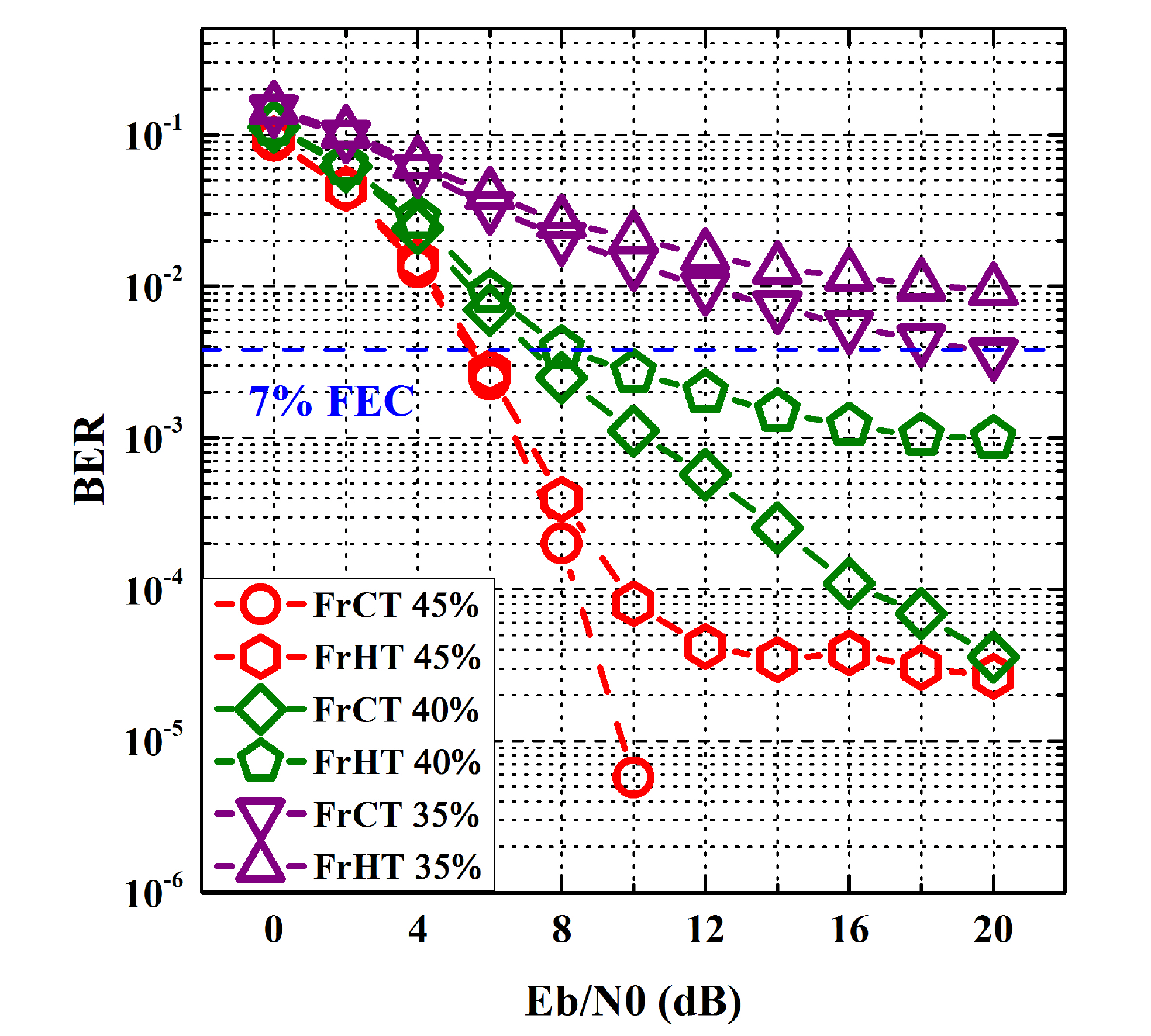}
\caption{BER versus $E_{b}/N_{0}$ for FrCT-NOFDM and FrHT-NOFDM with different subcarrier spacing.}\label{fig:7}
\end{figure}

Figure \ref{fig:7} reveals BER versus $E_{b}/N_{0}$ for FrCT-NOFDM and FrHT-NOFDM with different subcarrier spacing. The iterative number of ID algorithm is set to 20. The subcarrier spacing is equal to $45\%$, $40\%$ and $35\%$ of the symbol rate per subcarrier when $\alpha$ in FrCT-NOFDM is set to 0.9, 0.8 and 0.7, respectively and $\alpha$ in FrHT-NOFDM is set to 0.45, 0.4 and 0.35, respectively. As discussed above, the ICI in FrCT-NOFDM should be smaller than that in FrHT-NOFDM. After ID algorithm, the residual ICI in FrCT-NOFDM is also smaller than that in FrHT-NOFDM. Therefore, the BER performance of FrCT-NOFDM is much better than that of FrHT-NOFDM. When the subcarrier spacing is set to $45\%$ of symbol rate per subcarrier, the BER of FrCT-NOFDM descends with the increase of $E_{b}/N_{0}$, but the BER of FrHT-NOFDM no longer decreases because the residual ICI begins to be the major distortion when the $E_{b}/N_{0}$ is larger than $10$ dB. For generating FTN NOFDM signal, FrCT is a much better choice than FrHT.

\begin{figure}[!t]
\centering
\includegraphics[width=3in]{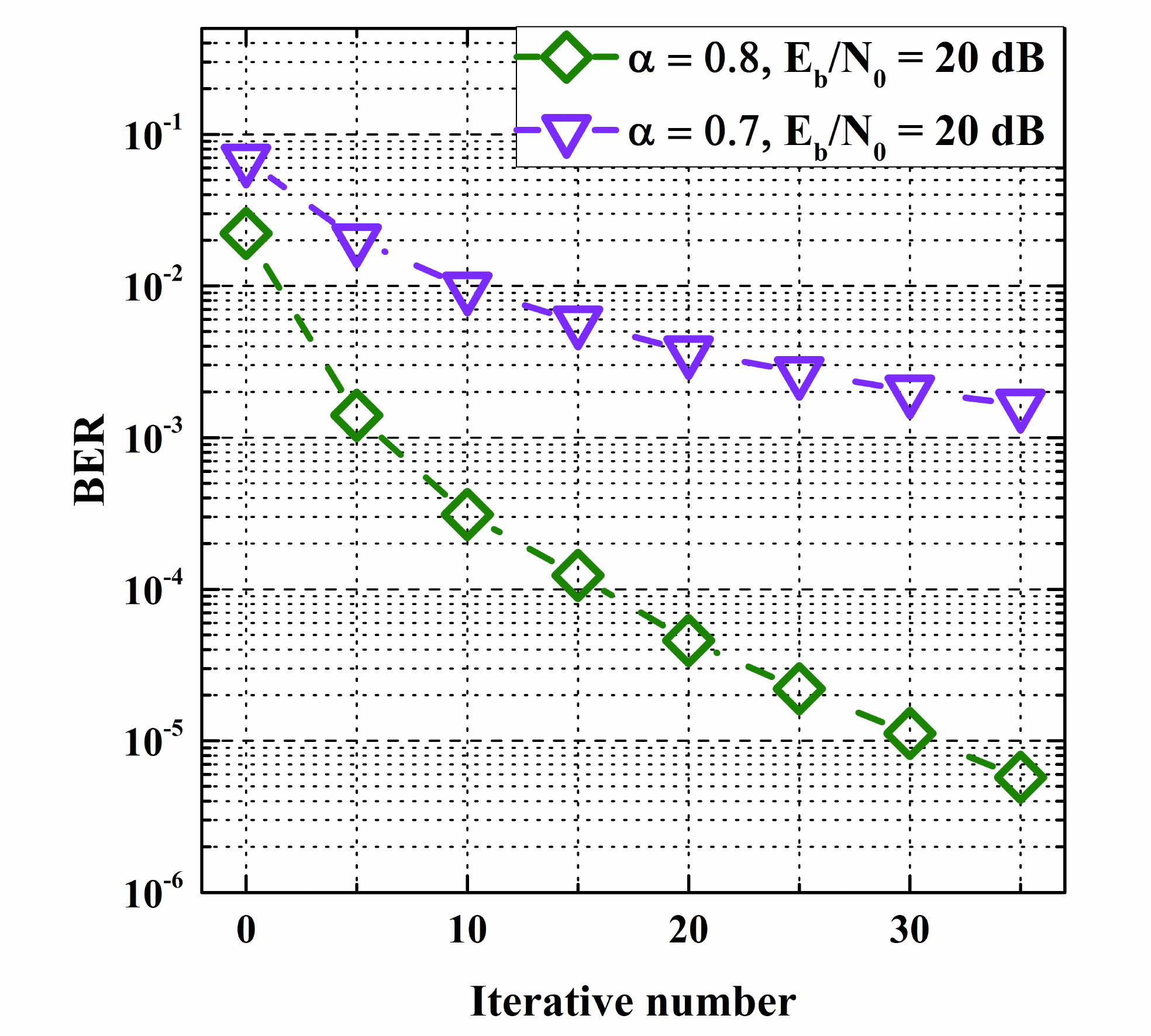}
\caption{BER against iterative number of the ID algorithm.}\label{fig:8}
\end{figure}

Figure \ref{fig:8} shows the BER against iterative number of ID algorithm for FrCT-NOFDM when $E_b/N_0$ is set to 20 dB. When $\alpha$ is set to $0.8$, BER performance can be improved by the increase of iterative number. Therefore, it has the potential to further eliminate the ICI in FrCT-NOFDM with $\alpha$ of 0.8. However, when $\alpha$ is set to $0.7$, the increase of iterative number cannot significantly improve the BER performance. This is because the ICI is too large to be completely eliminated by ID algorithm. Therefore, it needs a more effective algorithm to compensate the interference. The ID-fixed sphere decoder (ID-FSD) algorithm has a better performance than ID algorithm, which combines the ID and FSD algorithms\cite{Darwazeh:2013CL}. Furthermore, the channel coding is effective for resisting the ICI and has been investigated for improving the BER performance of NOFDM \cite{Anderson:2013IEEE, Xu:2014CL}.

The simulation results verify the feasibility of the FrCT-NOFDM. When $\alpha$ is larger than a certain value (this value may be no more than 0.8), ICI can be effectively eliminated and FrCT-NOFDM can achieve almost the same BER performance compared to DCT-OFDM. This certain value is similar to the Mazo limit for FTN-NOTDM systems\cite{Mazo:1975TBSTJ}. Therefore, when the $\alpha$ is greater than this certain value, the capacity of FrCT-NOFDM has the possibility to approach to the capacity limit.

\subsection*{Proof-of-concept experiment}
To verify the feasibility of the FrCT-NOFDM, we set up an experiment as shown in Figure \ref{fig:1}. At the transmitter end, the size of FrCT was set to $256$ and 2-PAM was modulated. Sixteen cyclic prefix samples were employed. One frame included $128$ FrCT-NOFDM symbols, $10$ training symbols and $1$ synchronization symbol. The digital signal was uploaded into an arbitrary waveform generator (Tektronix AWG7122C) operating at $10$ GS/s to realize digital-to-analog conversion. The resolution of arbitrary waveform generator was set to $8$ bits. The overall link rate was $10$ Gbit/s and the net bit rate was approximately $8.7$ Gbit/s ($1~\text{bit/sample}\times 10~\text{GS/s} \times 256/(256+16)\times 128/(128+10+1) \approx 8.7~\text{Gbit/s}$). An external cavity laser (ECL) with a linewidth of 100 kHz was used to generate the optical carrier. A Mach-Zehnder modulator (MZM) was used to modulate the optical carrier with the generated electrical signal. The $V_{\pi}$ of the MZM is about $1.5$ V and the bias voltage is set to about $1.5$ V.

The launch optical power was set to $3$ dBm. The length of standard signal mode fiber (SSMF) was $50$ km. Its total loss was approximately $10$ dB. A variable optical attenuator was employed to change the received optical power.

At the receiver end, the received optical signal can be converted into an electrical signal by the photodiode (Discovery DSC-R401HG). The electrical signal was then filtered by a low-pass filter with a 3-dB bandwidth of $10$ GHz. The filtered electrical signal was captured by a real-time digital phosphor oscilloscope (Tektronix DPO72004C) operating at $50$ GS/s to implement analog-to-digital conversion. The generated digital signal was decoded by off-line processing in MATLAB. The training symbols were used to estimate the channel characteristics for intra-symbol frequency-domain averaging (ISFA) algorithm and frequency-domain equalization \cite{Liu:2008OE}. After equalization, ID algorithm was employed to reduce the ICI.

\begin{figure}[!t]
\centering
\includegraphics[width=4in]{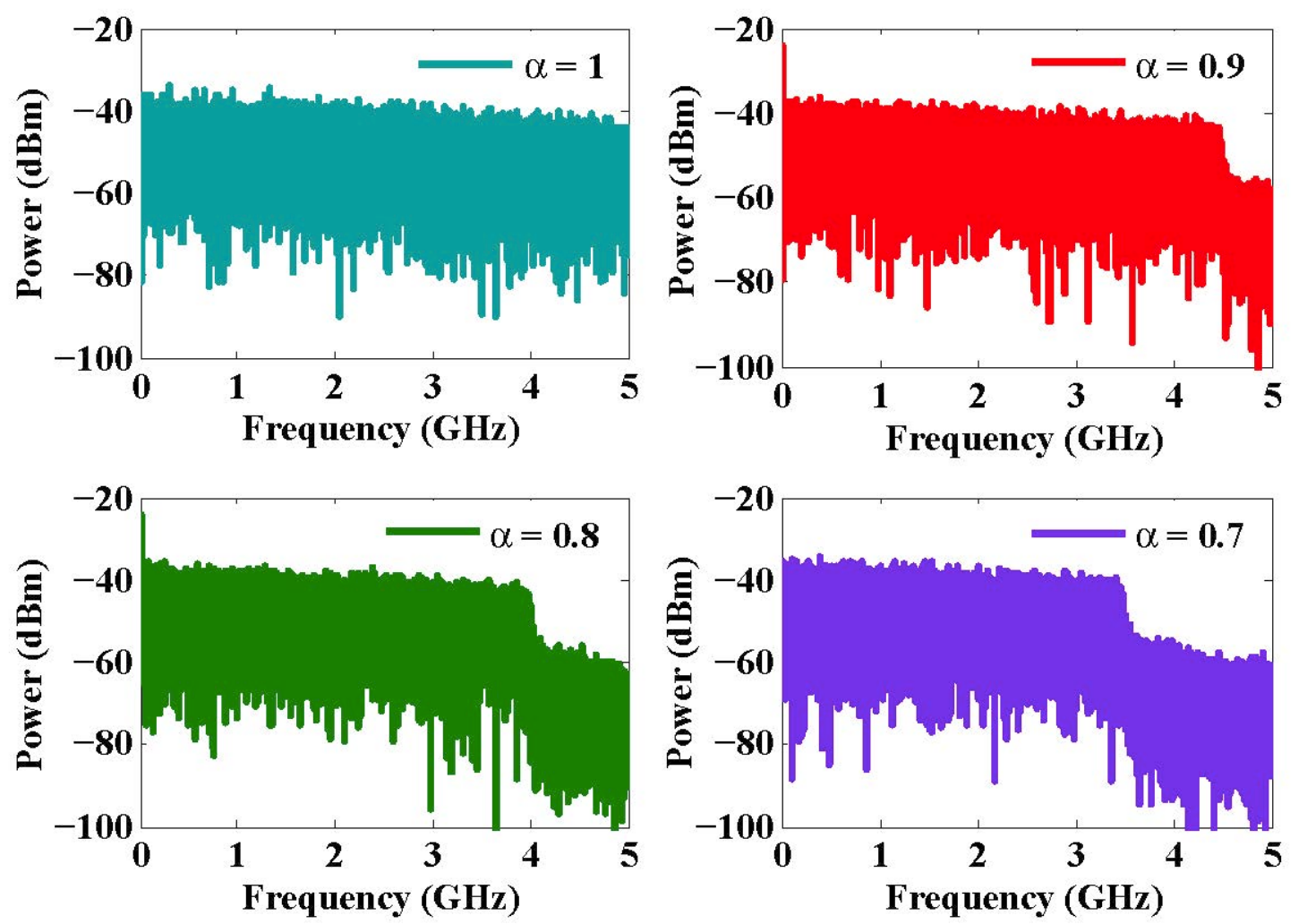}
\caption{Electrical spectra of DCT-OFDM (i.e., $\alpha = 1$) and FrCT-NOFDM signal.}
\label{fig:9}
\end{figure}

Figure \ref{fig:9} shows the electrical spectra of DCT-OFDM (i.e., $\alpha = 1$) and FrCT-NOFDM signal. The baseband bandwidth of DCT-OFDM is equal to $5$ GHz which is the Nyquist frequency of the signal with a $10$-GS/s sample rate. In FrCT-NOFDM, the baseband bandwidth is compressed to $4.5$, $4$ and $3.5$ GHz and the corresponding Nyquist rate is $9$, $8$ and $7$ Gbit/s when $\alpha$ is set to $0.9$, $0.8$ and $0.7$, respectively. In the experiment, the link rate of FrCT-NOFDM is 10 Gbit/s, which is faster than the corresponding Nyquist rate. Therefore, the FrCT-NOFDM signal is a kind of FTN signal. When $\alpha$ is set to $0.8$, $20\%$ baseband bandwidth saving can be obtained and the transmission rate is about $25\%$ faster than the Nyquist rate. These results verify the above theory analysis.

\begin{figure}[!t]
\centering
\includegraphics[width=3in]{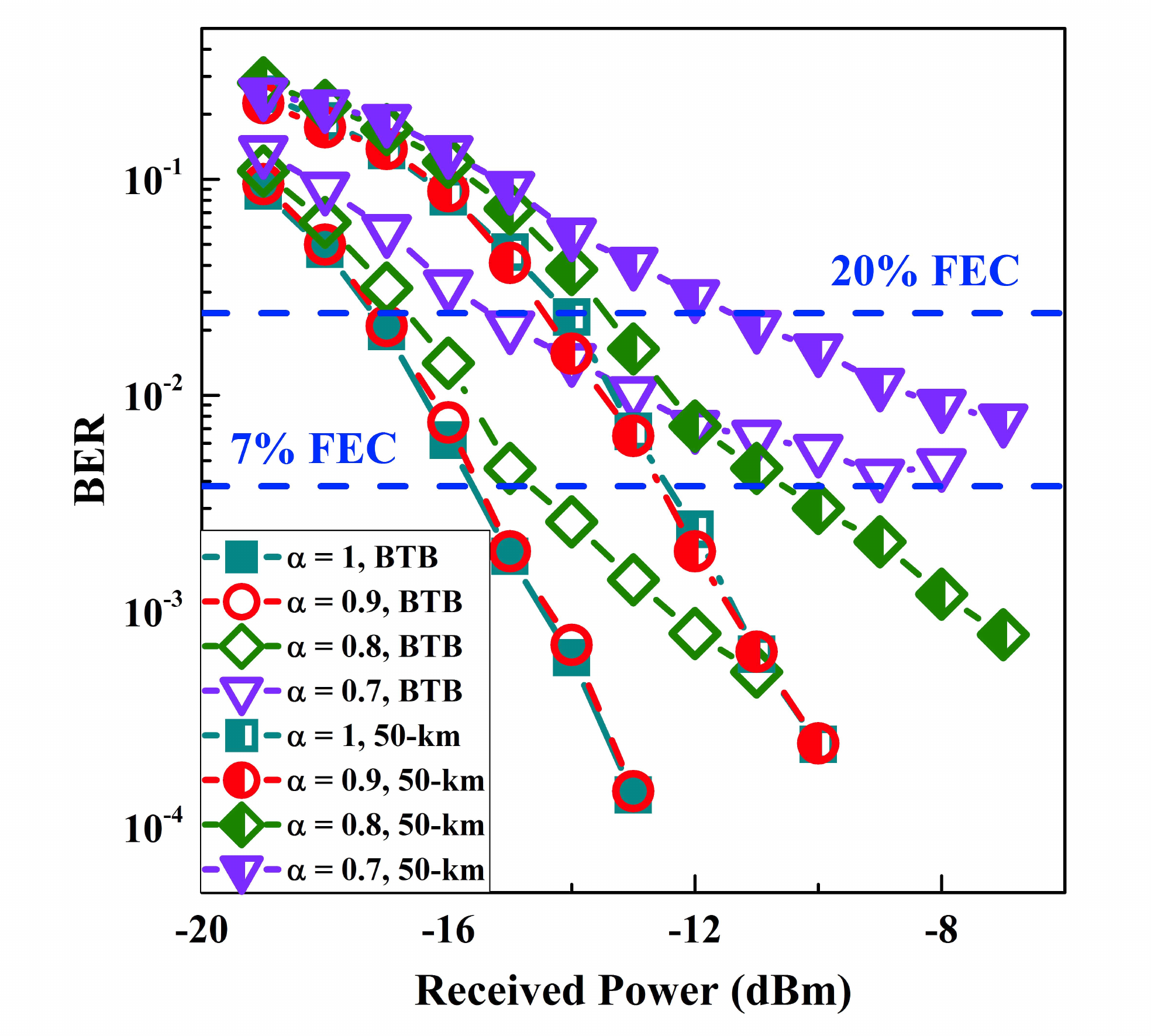}
\caption{BER curves for DCT-OFDM (i.e., $\alpha = 1$) and FrCT-NOFDM after back-to-back (BTB) and 50 km SSMF transmission.}
\label{fig:10}
\end{figure}

Figure \ref{fig:10} depicts the BER curves for DCT-OFDM (i.e., $\alpha = 1$) and FrCT-NOFDM after back-to-back (BTB) and 50-km SSMF transmission. The iterative number of the ID algorithm is set to $20$. FrCT-NOFDM with $\alpha$ of $0.9$ has the same BER performance with DCT-OFDM. After 50-km SSMF transmission, the required received power at the $7\%$ FEC limit was measured to be approximately $-12$ dBm for FrCT-NOFDM with $\alpha$ of 0.9. Compared to BTB transmission, the  power penalty for 50-km SSMF transmission is about $3$ dB. When $\alpha$ is set to $0.8$, the ICI degrades the BER performance. After 50-km SSMF transmission, the required received power at the $7\%$ FEC limit was measured to be approximately $-10$ dBm for FrCT-NOFDM with $\alpha$ of 0.8. Compared to DCT-OFDM, the power penalty for FrCT-NOFDM with $\alpha$ of $0.8$ is about $2$ dB at the $7\%$ FEC limit. When $\alpha$ is set to $0.7$, the ICI severely degrades the BER performance, the BER can only achieve the $20\%$ FEC limit.

\begin{figure}[!t]
\centering
\includegraphics[width=3in]{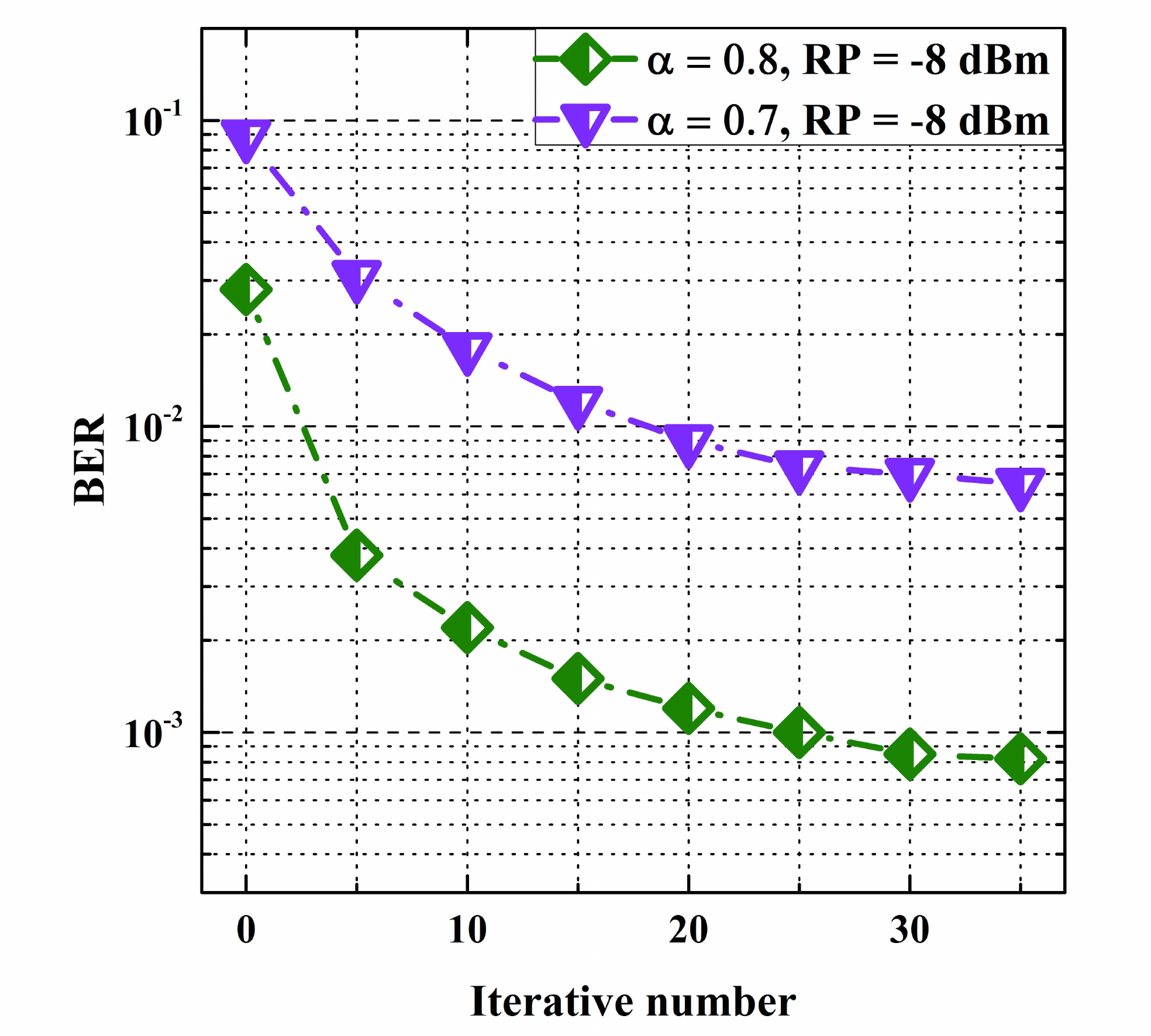}
\caption{BER versus iterative number of the ID algorithm after 50-km SSMF transmission.}\label{fig:11}
\end{figure}

Figure \ref{fig:11} depicts the BER against iterative number of the ID algorithm after 50-km SSMF transmission. The received power is set to $-8$ dBm. The $\alpha$ is set to $0.8$ and $0.7$, respectively. In theory, ID algorithm with more iterative number can eliminate more ICI. Therefore, the BER decreases with the increase of iterative number. However, when iterative number is larger than $20$, the increase of iterative number cannot significantly improve the BER performance. This may be because the other distortions such as chromatic dispersion (CD) mainly influence the BER performance when the ICI is small. The experiment results agree well with the simulation results. The feasibility of FrCT-NOFDM has been verified by both the experiment and simulation.

\section*{Discussion}
The history of FTN signal began with the paper of James Mazo in $1975$\cite{Mazo:1975TBSTJ}, who investigated the time-domain binary sinc-pulse scheme. The FTN binary sinc-pulse scheme compresses the time period to obtain the transmission rate faster than Nyquist rate. The pulses are accelerated with the time acceleration factor $\tau$ and become no longer non-orthogonal, thus the FTN sinc-pulse signal can be considered as a kind of FTN NOTDM signal. Twenty-five percent more bits can be carried in the same bandwidth while $\tau$ is set to 0.8, now known as the Mazo limit. In this paper, we demonstrate the principle of FTN NOFDM by taking FrCT-NOFDM for instance. The subcarrier spacing is less than $50\%$ of the symbol rate per subcarrier. When the bandwidth compression factor $\alpha$ is set to $0.8$, the subcarrier spacing is equal to $40\%$ of the symbol rate per subcarrier, thus $20\%$ baseband bandwidth saving can be obtained and the transmission rate is about $25\%$ faster than the Nyquist rate.

In 1940s, Shannon put forward the up limit of communication capacity for the Nyquist signal \cite{Shannon:1948TBSTJ, Shannon:1949PIRE}. When the ISI is effectively compensated, the FTN NOTDM signal can achieve a higher capacity limit than Nyquist signal \cite{Anderson:2013IEEE, Rusek:2009TIT}. To the best of our knowledge, we first give the mathematical expression for capacity limit of FTN NOFDM signal,
\begin{equation}\label{eq:eq15}
C \leq \frac{1}{\alpha}\times  Wlog_{2}\left(1+\frac{P_S}{P_N+P_{ICI}}\right).
\end{equation}
The capacity limit of FTN NOFDM signal is potentially higher than that of Nyquist signal when ICI is effectively compensated. In other word, the capacity of FTN-NOFDM signal is likely to approach the limit only when ICI has been effectively eliminated. Therefore, how to eliminate the ICI is crucial for FTN-NOFDM.

As verified in literatures \cite{Mazo:1975TBSTJ, Anderson:2013IEEE}, when the $\alpha$ is set to the value between $1$ and $0.8$, there is not obvious deterioration in BER performance by employing the optimal detection to compensate the interference. In this paper, we demonstrate the simulations and experiments of FrCT-NOFDM system, which verify that ICI can be effectively eliminated when the $\alpha$ is set to the value between 1 and 0.8. As a result, the capacity limit of FrCT-NOFDM signal is potentially higher than that of the Nyquist signal when the $\alpha$ is set to the value between 1 and 0.8. In the further work, more effective algorithm can be employed to eliminate the ICI and higher-order $M$-PAM constellation can be investigated to obtain higher spectral efficiency. FrCT-NOFDM can be potentially used in next-generation high-capacity wireless and optical communications.

\renewcommand{\algorithmicrequire}{ \textbf{Input:}}
\renewcommand{\algorithmicensure}{ \textbf{Output:}}
\begin{algorithm}[!t]
\caption{ID algorithm for $2$-PAM constellation.}
\label{alg:1}
\begin{algorithmic}[1]
\Require~~ Received symbol : $\textbf{R}$; Compression factor : $\alpha$; Iterative number : $I$
\Ensure ~~ Recovered symbol : $\textbf{S}$
\For{$l=0$; $l<N$; $l++$} \Comment{Calculating correlation matrix $\textbf{C}$}
\For{$m=0$; $m<N$; $m++$}
\State $C_{l,~m} = \frac{2}{N}\sum_{n=0}^{N-1}W_{l} \text{cos}\left(\frac{\alpha \pi l(2n+1)}{2N}\right)\cdot W_{m} \text{cos}\left(\frac{\alpha \pi (2n+1)m}{2N}\right)$.
\EndFor
\EndFor
\State Initialization : $\textbf{S}_0 = 0$, $d = 1$ \Comment{Iterative operation}
\For{$i = 1$; $i\leq I$; $i++$}
\State $\textbf{S}_i = \textbf{R} - (\textbf{C}-\textbf{e})\textbf{S}_{i-1}$
\If{$\textbf{S}_i > d$} \Comment{Constellation mapping}
\State $\textbf{S}_i = 1$
\ElsIf{$\textbf{S}_i < -d$}
\State $\textbf{S}_i = -1$
\Else
\State $\textbf{S}_i = \textbf{S}_i$
\EndIf
\State $d = 1-i/I$ \Comment{Updating $d$}
\EndFor
\State $\textbf{S} = \textbf{S}_I$
\State Return $\textbf{S}$
\end{algorithmic}
\end{algorithm}

\section*{Methods}
The simulations and off-line processing in the experiment were both implemented by MATLAB. The encoding and decoding were shown in Figure \ref{fig:1}. In the simulation, the AWGN channel was employed. The channel equalization is not required for the AWGN channel. In the experiment, the training symbols were used to estimate the channel characteristics. Meanwhile, the ISFA algorithm can improve the performance of channel estimation \cite{Liu:2008OE}. The frequency-domain equalization can compensate the channel distortion by using the estimated channel characteristics.

In NOFDM system, ID algorithm is employed to eliminate the ICI, which is critical to improve the BER performance. The ID algorithm for $2$-PAM constellation is shown in Algorithm \ref{alg:1}. Compared to two-dimensional constellation (i.e., $M$-QAM), the constellation mapping in ID algorithm for one-dimensional constellation (i.e., $M$-PAM) is more simple and accurate.

\section*{Acknowledgements (not compulsory)}
This work was supported in part by National Natural Science Foundation of China (61427813, 61331010, 61271192); BUPT Excellent Ph.D. Students Foundation; China Scholarship Council Foundation.

\section*{Author contributions statement}
J. Z. and Y. Q. derived the theoretical results, J. Z. and Y. Q. conceived the simulation and experiment, J. Z., M. G., and X. T. conducted the simulation and experiment, J. Z., Z. Y., Q. C., and Q. W. analyzed the results. All authors reviewed the manuscript.

\section*{Additional information}
\textbf{Competing financial interests:} The authors declare no competing financial interests.

%\end{linenumbers}
\end{document}